# Data Analysis: Communicating with Offshore Vendors using Instant Messaging Services

Jongkil Jay Jeong

## I. Background & Approach

In empirical studies, it is important to apply proper data analysis techniques so that useful information and correct conclusions can be derived from it. In particular, qualitative studies in the social science field require a large amount of data to be collected and processed based on the actions and events in our daily social life. Therefore, without the correct interpretation and simplification of this vast amount of data, it is difficult to really make sense as to what the true meaning of this data is referring to. In this regard, researchers should pay special attention to process the collected data by inspecting, cleaning, transforming and modelling it to ensure the correct interpretation of the empirical results.

The purpose of this study is to find whether the choice of correct analytic process is effective to derive a meaningful and correct conclusion from the vast amount of information. For this purpose, I designed an analytic framework to investigate the importance of effective communication on the success of IT business. The analytic process used for this project is the Grounded Theory approach so that the data collected could be developed into a more refined theoretical framework. Grounded theory is a widely used approach in qualitative research, and is defined as "a qualitative research method that uses a systematic set of procedures to develop an inductively derived theory about a phenomenon" (Strauss & Corbin 1990).

In this study, Grounded Theory approach is applied to a collected communication data between Client and Offshore Supplier in IT business. For data, conversation logs* are obtained over a 6 month period between November 2010 to May 2011 from an instant messaging program, between an IT Consultant ("Client") from Melbourne, Australia and the team lead from his offshore development office from New Delhi, India ("Offshore Supplier").

The two have been working on multiple projects with each other for approximately 25 months, and communicates with each other on a day to day basis via Email, Phone Calls and Instant Messaging online.

As the amount of data in these conversation logs were too broad for the scope of this project, excerpts of conversations via Instant Messaging as provided in Appendix B were chosen based on four main topics which was identified to be the most frequent type of communication between the Client and Offshore Supplier.

The type of communication can be classified by the following four topics:

   a. The Client briefing the Offshore Supplier on a new project.
   b. The Client and Offshore Supplier working through specific issues (conflict management).
   c. The Client and Offshore Supplier discussing matters related to existing projects
   d. The Client and Offshore supplier discussing matters related to payment & accounts

*The Conversation logs were edited and formatted so that the names of the project and people involved were kept anonymous*

## II. Analytic Process & Strategy

In order for the theory to be derived, data was put through a three step process of transformation based on Neuman's (2006) coding & concept formation techniques. They are open coding, axial coding and selective coding. The results of Neuman's (2006) coding & concept formation techniques are as the following.

a. **Open Coding**: The purpose of open coding is to examine the data collected to condense them into preliminary analytic categories or codes. This allows specific themes based on key events, critical terms and central people to be categorised and identified in a manner that allows for future evaluation of the data in more detail in latter phases of the data analysis process (Neuman 2006).

During the open coding process, the following concepts were identified:

a) Frequent grammatical errors & spelling mistakes across the board did not seem to impede communication between the client & offshore provider.
b) Acronyms tend to be used more frequently when conversation was related to new & existing projects.
c) A conversation tends to stay focused on one particular topic and does not change until the point surrounding the topic has been addressed.
d) Sentences tend to be short & split across multiple lines instead of full sentences.
e) Timeliness of delivery was emphasised throughout all the conversations and is perceived as the biggest issue & concern by the client.
f) All conversations lasted less than 30 minutes, and were direct in nature.
g) Client's demands appear to have more emphasis and priority compared to the Offshore Provider's requests.
h) Language used by the offshore provider has elements of honorific characteristics.

The coding structure for three of these concepts is presented below as an example (A complete list of the coding for each concept is provided in Appendix A):

**Concept A**

**Label:** Grammatical Errors & Spelling Mistakes

**Definition:** Frequent grammatical errors and spelling mistakes are observed throughout all the conversations between the Client & Offshore Provider, but both parties seem to understand in detail what the conversation is about.

**Flag:** Any spelling mistakes, grammatical inaccuracy or incorrect use of vocabulary may all be instances of this label being flagged.

**Qualifications:** Only grammatical mistakes & spelling / vocabulary errors which may significantly hinder communication will be considered. Any minor mistakes to spelling that arise from typos will be discarded.

**Example:** In regards to the client wanting an explanation for some of the projects falling behind schedule, the Offshore provider responds with: *"No I m not givigin any any excuse ..I have the capability to show you a somewhat running system by Friday ..but my only concerns is the answer of my basis question that I aksed"*. Furthermore, a similar observation can be made when clarifying payment terms by the Offshore Provider: *"there reaming amt can come later"*.

**Concept B**

**Label:** Use of acronyms & abbreviations to communicate

**Definition:** Frequent use of acronyms by both parties to explain technical specifications is understood and acknowledged by both the Client & Offshore Provider.

**Flag:** Use of acronyms to describe specific technical functions / features, and abbreviations which are used when communicating with each other are all flags when this is occurring.

**Qualifications:**  Only technical acronyms & abbreviations will be considered. Furthermore, any internet *slang* (e.g. Lol etc.) will also be included.

**Example:** From the conversation logs related to the new project briefing, the offshore provider asks the client: *"now tell me will the main site be CMS or just static",* in which the client responds: *"the main site will be CMS based."* Another example is when the client & offshore provider are discussing about implementing an "API" to an existing project, where both parties understood in detail what this was referring to.

### Concept C

**Label:** Focused & Concentrated Conversations

**Definition:** Conversations tend to be focused on one particular subject and/or topic for the duration of the chat between the client & offshore provider, and the language used during this conversation tend to match the nature of the topic at hand.

**Flag:**  Discussions surrounding one particular project for the duration of the chat session may be an example of this occurring.

**Qualifications:** Reference to specific projects or topics changing in between sentences communicated via the instant messaging platform will be included as part of this concept.

**Example:** As per the topics defined in the background section, the data collected shows that there is a particular topic that the client & offshore provider want to discuss. For example, the conversation related to an existing project sticks to details around "Project A" throughout the duration of the chat.

b. **Axial Coding**: The purpose of axial coding is to re-examine and review the initial coded concepts from the open coding process to see whether or not these concepts can be organised and linked together or if new concepts and ideas can be derived from the concepts gathered in the open coding process (Neuman 2006).

During the examination of the concepts through axial coding, the following refined & grouped themes were identified:

a) The use of proper grammar and vocabulary is not necessarily required in order for the client and offshore provider to understand each other, as long as the language used between the client & offshore supplier was within the context of what needs to be communicated (Concepts A, B, C combined).
b) Conversations were difficult to understand if the observer did not have any prior background about the topic which was being discussed (Concepts B & D combined).
c) Conversations tend to be short and focused, and did not deviate away from the major topic of discussion (Concept C & F combined)
d) Cultural differences on how work was done between the client & offshore provider were evident (Concept E, G, H combined).

After the grouping of concepts, the newly derived themes were examined again to see if there were connections to any major topic of interest (Neuman 2006). Certain key topics did emerge from the four themes outlined above, and these were as follows:

- Method of Communication - **Language**
- Knowledge & Mutual understanding - **Trust**
- Differences in how work was conducted – **Cultural Differences**

c.  **Selective Coding:** The final process in coding qualitative data is Selective coding, where previous codes are examined and data is selected, so that the conceptual coding categories that were developed in the previous phases can be supported (Neuman 2006).

Through the analysis of data and concepts from the previous sections, a core theme was identified and became evident:

"Errors or mistakes associated with the grammar, lexical resource & sentence structures surrounding conversations via instant messaging services do not impede the level of understanding as long as there is a high level of trust and mutual understanding between the client & offshore provider and the language used for the conversation is appropriate for discussing the topic at hand."

## III. Interpretation of Data

Based on the analysis of the data, it is apparent that communication was still possible between the client & offshore provider despite the high amount of errors and mistakes associated with the spelling & grammar. This was in part due to the fact that as long as the language used between the parties was within the context of what needed to be communicated, any issues that may arise from grammatical and lexical mistakes could be rectified and understood.

A high level of fragmentation in the sentence structuring was also identified through the data analysis process but this also had minimal impact on the level of understanding between the two parties. This was due to the fact that despite the high level of fragmentation, conversations tend to be short and focused, and did not deviate away from the main discussion topic.

Furthermore, the data analysed also revealed that the language and vocabulary used changed significantly based on the nature of the topic being discussed, and this was evident based on the different acronyms and abbreviations that were observed throughout the data.

Finally, cultural differences could also be observed throughout the data and this led to different language being used to reference one another, as well as observations that the primary concern of the client & offshore provider were different – the client emphasized adherence to schedule, while the service provider emphasized gathering additional information.

In conclusion, this means that prior knowledge and mutual understanding between the two service providers, acknowledging that cultural differences meant that how work was conducted may be different as well as agreeing to a specific method of communication via a common language would help enhance the level of understanding and communication between clients & offshore service providers.

## IV. Research Implications, Limitations and Suggestions for Future Study

The conversation logs analysed has provided help in understanding in more detail the key characteristics surrounding communication between clients and offshore providers.

However, it is also acknowledged that there are certain limitations to the data. As the findings from the data was derived from only one case study based on a particular offshore provider and client, additional research with more case studies containing different backgrounds, experience and environments is suggested so that the findings made through this data set can be verified to be conclusions which can be observed in generic cases of communication with offshore providers via instant messaging.

Furthermore, the data does not show how much previous communication surrounding the topics discussed in the chat logs occurred through other methods such as email, voice conversations or formal documentations.

It is recommended that future scholars researching within this field, investigate the preferred method of communication by parties, and how much time is spent percentage wise using a particular method communicating and interacting with their offshore teams on a given project. In addition, it may be worthwhile understanding the characteristics and strengths / weaknesses of each method of communication which may be valuable information for current industry practitioners working within the outsourcing / offshoring field.

## V. Conclusion

By undertaking the grounded theory approach, the conclusion reached was that the language used in instant messaging environments between clients & offshore providers was highly fragmented and broken, but both the client and offshore provider did not seemed to be impacted by these anomalies.

The reasoning behind this was that mutual understanding & knowledge between the two parties, and choosing a specific language that is relevant to the conversation at hand was enough for both sides to make sense of what was being said.

Based on this finding, I encourage future scholars and industry practitioners to continue on carrying out work which examines the relationship between language, trust, communication and culture in reference to ICT offshoring.

## Appendix A: Data Coding

### A-1. Concept A

**Label:** Grammatical Errors & Spelling Mistakes

**Definition:** Despite frequent grammatical errors and spelling mistakes being observed throughout all the conversations between the Client & Offshore Provider, communication between the two parties seem to not have been affected.

**Flag:** Any spelling mistakes, grammatical inaccuracy or incorrect use of vocabulary may all be examples of when this is occurring.

**Qualifications:** Only grammatical mistakes & spelling / vocabulary errors which may significantly hinder communication will be considered. Any minor mistakes to spelling that arise from typos will be discarded.

**Example:** In regards to the client wanting an explanation for some of the projects falling behind schedule, the Offshore provider responds with: *"No I m not givigin any any excuse ..I have the capability to show you a somewhat running system by Friday ..but my only concenrs is the answer of my basis question that I aksed"*. Furthermore, a similar observation can be made when clarifying payment terms by the Offshore Provider: *"there reaming amt can come later"*.

### A-2. Concept B

**Label:** Use of acronyms & abbreviations to communicate

**Definition:** Frequent use of acronyms by both parties to explain technical specifications is understood and acknowledged by both the Client & Offshore Provider.

**Flag:** Use of acronyms to describe specific technical functions / features, and abbreviations which are used when communicating with each other are all flags when this is occurring.

**Qualifications:** Only technical acronyms & abbreviations will be considered. Furthermore, any internet *slang* (e.g. Lol etc.) will also be included.

**Example:** From the conversation logs related to the new project briefing, the offshore provider asks the client: *"now tell me will the main site be CMS or just static"*, in which the client responds: *"the main site will be CMS based."* Another example is when the client & offshore provider are discussing about implementing an "API" to an existing project, where both parties understood in detail what this was referring to.

### A-3. Concept C

**Label:** Focused & Concentrated Conversations

**Definition:** Conversations tend to be focused on one particular subject and/or topic for the duration of the chat between the client & offshore provider, and the language used during this conversation tend to match the nature of the topic at hand.

**Flag:** Discussions surrounding one particular project for the duration of the chat session may be an example of this occurring.

**Qualifications:** Reference to specific projects or topics changing in between sentences communicated via the instant messaging platform will be included as part of this concept.

**Example:** As per the topics defined in the background section, the data collected shows that there is a particular topic that the client & offshore provider want to discuss. For example, the conversation related to an existing project sticks to details around "Project A" throughout the duration of the chat.

### A-4. Concept D

**Label:** Fragmented & Split sentences

**Definition:** Sentences from the data tend to be split and fragmented across multiple lines, instead of one full sentence structure.

**Flag:** An example may be when a sentence referring to one specific issue is split across multiple lines.

**Qualifications:** If a sentence related to a particular subject or a topic is broken into multiple chat lines instead of one. Any breaks in the conversation due to clarifying a point or rectifying a term will not be included.

**Example:** The client is trying to explain to the offshore provider to continue on with the development of the work, but hold on the API implementation until the launch of the final site:

> *[23/02/2011 4:42:34 p.m.] Client: don't stop the dev work though*

> *[23/02/2011 4:42:45 p.m.] Client: at the very least, what we will do is implement the API after the launch of the site*

> *[23/02/2011 4:42:48 p.m.] Client: and charge extra for the work.*

### A-5. Concept E

**Label:** Timely delivery of projects

**Definition:** There is significant emphasis by the client on adhering to the project schedule and keeping to a specific timeframe.

**Flag:** Any reference to date, time or timeframe for a particular project.

**Qualifications:** Reference to date, time or timeframe in the context of a specific delivery or schedule of a project will be qualified. Any reference to time in regards to a personal question (e.g. What is the time right now?) is not considered.

**Example:** The client seems to put pressure on the offshore provider referring to the fact that certain projects are running behind schedule. An example of this is when the client says "Well, a week for 2 mock designs is plenty of time tbh." Which the offshore responds by saying: "I reviewed the API yesterday ...and will discuss them with developers today ....hopefully by tomorrow I will present you few modules of Project A working. Yesterday and today we have kept as delivery days for Project A".

### A-6. Concept F

**Label:** Length & Directness of conversation

**Definition:** The data shows that conversations on topics are in most cases very short, and direct in nature.

**Flag:** Observing the time of message sent and reply received between the offshore provider & client, and then measuring how long they converse about the topic before moving to the next.

**Qualifications:** Conversations must be in scope of a particular issue, and any data that was observed which was making multiple references to different projects was not considered.

**Example:** Despite calculating quite a number of projects for payment, the entire conversation only took 22 minutes (2:59pm – 3:22pm) to finalise.

### A-7. Concept G

**Label:** Client & Supplier Demands & Requests

**Definition:** The client tends to focus on clarification & demands, whereas the offshore provider tends to make more requests.

**Flag:** An example of this is when the client asks for further clarification or information from the offshore provider.

**Qualifications:** Only genuine requests or demands will be taken into account. Any rhetorical questions observed will not be considered.

**Example:** A client asks the offshore provider for clarification on a particular project by saying: *"Project A: Still need your comments re the API."* which is more of a demand / request for clarification on a particular issue, whereas the offshore provider makes requests: *"can you please arrange the softcopy for the logo"*

### A-8. Concept H

**Label:** Language used to reference each other is different

**Definition:** There is a difference as to how the client is referred to by the offshore provider, and vice versa when looking through the data.

**Flag:** Reference to the client was made by the offshore provider as "sir", whereas the client referred the offshore provider by his/her name.

**Qualifications:** Any use of language that has honorific elements will be deemed to be qualified for this concept.

**Example:** Client refers to his offshore provider by name *"it needs to be done Offshore Supplier (Name)"*. In contrast, the offshore provider refers the client as "sir" on a frequent basis.

## Appendix B: Instant Messaging Conversation Logs

**B-1. Conversation Related to New Project Brief**

[20/04/2011 5:46:14 p.m.] Client:  New Project - I have now clarified the details for this, and will brief you via email shortly

[20/04/2011 5:46:22 p.m.] Client: Essentially, it will be 6 websites in total

[20/04/2011 5:46:25 p.m.] Offshore Supplier: ok Sir

[20/04/2011 5:46:25 p.m.] Client: 1 Main site / 5 sub-sites

[20/04/2011 5:46:27 p.m.] Offshore Supplier: great

[20/04/2011 5:46:31 p.m.] Client: nothing complex.

[20/04/2011 5:46:36 p.m.] Client: just basic CMS websites

[20/04/2011 5:46:45 p.m.] Client: http://www.projectEwebsite.com

[20/04/2011 5:46:59 p.m.] Offshore Supplier: k

[20/04/2011 5:47:05 p.m.] *** Client sent Project E Design Brief.pdf ***

[20/04/2011 5:47:17 p.m.] Client: They tried to do full flash: http://www.webmarketingworks.com.au/SFGDemo/Flash3/ but it didnt work out

[20/04/2011 5:47:36 p.m.] Client: Anyways, the main site will be a list of all 5 sites with products & services + links to the individual sites

[20/04/2011 5:47:58 p.m.] Client: the 5 sub sites will be basic 7 - 10 page CMS based static websites (we will not be using flash for this project) on Wordpress

[20/04/2011 5:49:15 p.m.] Offshore Supplier: ok So lets put it this way - that there will be one parent site and 5 child sites [With CMS]

[20/04/2011 5:49:19 p.m.] Client: yes

[20/04/2011 5:49:28 p.m.] *** Client sent Project Logo - Centered 630.png ***

[20/04/2011 5:49:31 p.m.] Offshore Supplier: now tell me will the main site be CMS or just static

[20/04/2011 5:49:48 p.m.] Client: the main site will be CMS based

[20/04/2011 5:49:53 p.m.] Client: but will be design focused

[20/04/2011 5:49:59 p.m.] Client: not much text - a lot of images

[20/04/2011 5:50:05 p.m.] Client: it will be used as a leadin site for the other 5

[20/04/2011 5:50:18 p.m.] Offshore Supplier: ok

[20/04/2011 5:50:21 p.m.] Offshore Supplier: cool

[20/04/2011 5:51:51 p.m.] Offshore Supplier: can you please arrange the softcopy for the logo

[20/04/2011 5:52:03 p.m.] Offshore Supplier: 2> How many desing do we have to make

[20/04/2011 5:52:11 p.m.] Offshore Supplier: 1-2 .6?

[20/04/2011 5:52:11 p.m.] Client: Looks like the main site will be a "landing page" site - design intenstive and the overall purpose is to lead the visitor the the 5 sub sites, based on what service they require

The 5 subsites will be basic CMS based websites which will share a simular look and feel

[20/04/2011 5:52:30 p.m.] Client: the 5 subsite designs will be all very simular

[20/04/2011 5:52:37 p.m.] Client: just different logo + content

[20/04/2011 5:52:49 p.m.] Offshore Supplier: same tempelate?

[20/04/2011 5:52:52 p.m.] Client: yes

[20/04/2011 5:54:56 p.m.] Offshore Supplier: ok

## B-2. Conversation Related to Existing Project

[23/02/2011 4:40:03 p.m.] Client: Project A: Still need your comments re the API.

[23/02/2011 4:40:07 p.m.] Offshore Supplier: yes

[23/02/2011 4:40:16 p.m.] Client: Well, a week for 2 mock designs is plenty of time tbh.

[23/02/2011 4:40:51 p.m.] Offshore Supplier: I reviewed the API yesterday ...and will discuss them with developers today ....hopefully by tomorrow I will present you few modules of Project A working

 [23/02/2011 4:41:08 p.m.] Offshore Supplier: yesterday and today we have kept as delivery days for Project A

[23/02/2011 4:41:16 p.m.] Client: just let me know

a. How long it will take

b. Is it going to cause any problems

c. What are some of the questions surrounding the implementation

[23/02/2011 4:41:26 p.m.] Offshore Supplier: ok sure!

[23/02/2011 4:41:53 p.m.] Offshore Supplier: are you ok if I revert you on Project A after todays/tomorrows delivery?

[23/02/2011 4:42:23 p.m.] Client: yes, Project A can wait

[23/02/2011 4:42:29 p.m.] Offshore Supplier: ok

[23/02/2011 4:42:34 p.m.] Client: don't stop the dev work though

[23/02/2011 4:42:45 p.m.] Client: at the very least, what we will do is implement the API after the launch of the site

[23/02/2011 4:42:48 p.m.] Client: and charge extra for the work.

[23/02/2011 4:43:20 p.m.] Offshore Supplier: yes ..i don't want to nag the team for API as I'm pushing for Project A delivery

[23/02/2011 4:43:30 p.m.] Offshore Supplier: btw any update from N

[23/02/2011 4:43:31 p.m.] Offshore Supplier: ?

[23/02/2011 4:44:26 p.m.] Client: yes, she is working on the site menu

[23/02/2011 4:44:30 p.m.] Offshore Supplier: ok

[23/02/2011 4:44:42 p.m.] Client: I also received a backup copy of the site

[23/02/2011 4:44:44 p.m.] Offshore Supplier: will be good if she sends it by Friday

[23/02/2011 4:44:45 p.m.] Client: via DVD

[23/02/2011 4:44:49 p.m.] Offshore Supplier: oh great!

[23/02/2011 4:44:55 p.m.] Client: and I am also getting the FTP details for you as we speak

[23/02/2011 4:44:56 p.m.] Offshore Supplier: when are they going down?

[23/02/2011 4:45:29 p.m.] Offshore Supplier: I really want the articles section of Project A site to be complete before they go down

[23/02/2011 4:46:58 p.m.] Offshore Supplier: Also, I need some more time on Project B as the contact us form on all the pages has messed up ...though its a small but stuck the thing ....will sort out everything by Friday

[23/02/2011 4:47:15 p.m.] Client: ok

[23/02/2011 4:47:29 p.m.] Client: please make sure you manage the time around delivery of projects

[23/02/2011 4:47:44 p.m.] Client: I urge you. It's very important now as we get more projects on board.

[23/02/2011 4:49:33 p.m.] Offshore Supplier: Everything will be on track tody

[23/02/2011 4:49:36 p.m.] Offshore Supplier: today*

[23/02/2011 4:49:46 p.m.] Client: ok

### B-3. Conversation related to addressing issues (delay in project)

[31/03/2011 9:07:37 a.m.] Client: it needs to be done Offshore Supplier.

[31/03/2011 9:07:40 a.m.] Client: We have had 3 weeks now

[31/03/2011 9:07:47 a.m.] Client: I have not seen one update for Project B since then.

[31/03/2011 9:08:22 a.m.] Client: Please do not use Project A as an excuse - you have had nearly 5 months for that project.

[31/03/2011 9:09:21 a.m.] Offshore Supplier: No I m not givigin any any excuse ..I have the capability to show you a somewhat running system by Friday ..but my only concenrs is the answer of my basis question that I aksed

[31/03/2011 9:09:25 a.m.] Offshore Supplier: you in my last email

[31/03/2011 9:09:34 a.m.] Client: yes - and I asked you to kindly put it in a spreadsheet

[31/03/2011 9:09:40 a.m.] Client: and send it because there are questions everywhere.

[31/03/2011 9:09:51 a.m.] Offshore Supplier: get me those anseres today and I iwll show you a good progress by Friday Promise

[31/03/2011 9:10:08 a.m.] Offshore Supplier: the excel sheet will reach you on some time from now

[31/03/2011 9:10:49 a.m.] Client: I also requested a project schedule (not an email) via Excel

[31/03/2011 9:10:54 a.m.] Client: this has not been done either.

[31/03/2011 9:11:37 a.m.] Offshore Supplier: ya..sorry ..I'm spendign more and more time with developers these days

[31/03/2011 9:11:46 a.m.] Offshore Supplier: sorry will send you excel

[31/03/2011 9:12:10 a.m.] Offshore Supplier: btw Traffic sites will be done before sat and Project E on monday

[31/03/2011 9:12:21 a.m.] Client: I have heard that so many times

[31/03/2011 9:12:24 a.m.] Client: I don't know what to believe Offshore Supplier

[31/03/2011 9:12:30 a.m.] Client: "It will be done by tomorrow"

[31/03/2011 9:12:33 a.m.] Client: "it will be done by this week"

[31/03/2011 9:12:46 a.m.] Offshore Supplier: http://www.xyz.com/

[31/03/2011 9:13:07 a.m.] Offshore Supplier: the developer is working with me at this point of time to get these traffic sites completed

[31/03/2011 9:13:44 a.m.] Offshore Supplier: once they are done we will get to the Project E ...Project E has only few prodcut pages left to be populated

[31/03/2011 9:14:39 a.m.] Offshore Supplier: Btw : past week the Project A team ran into an issue realted to role management which has eaten up all the time

[31/03/2011 9:15:00 a.m.] Offshore Supplier: of every body including me and html and wp developers

[31/03/2011 9:15:17 a.m.] Offshore Supplier: + the Cricket world cup has affected

[31/03/2011 9:15:37 a.m.] Offshore Supplier: that the honest reason of delays and issues from our end

[31/03/2011 9:16:04 a.m.] Client: yes

[31/03/2011 9:16:07 a.m.] Client: I thought that may be the case

[31/03/2011 9:16:12 a.m.] Client: FINALLY - you have been upfront and honest.

[31/03/2011 9:17:23 a.m.] Offshore Supplier: and today [wednesday ] we have India /pakistan semi final match - THIS WAS ONE OF THE MOST IMPORTANT EVENT IN THE INDIAN and PAKISTAN HISTORY of past 40 years

[31/03/2011 9:17:29 a.m.] Client: yes

[31/03/2011 9:17:33 a.m.] Client: I know India won

[31/03/2011 9:17:37 a.m.] Offshore Supplier: OH

[31/03/2011 9:17:40 a.m.] Offshore Supplier: GREAT

[31/03/2011 9:17:45 a.m.] Offshore Supplier: :)

 [31/03/2011 9:18:48 a.m.] Client: anyways

[31/03/2011 9:18:52 a.m.] Client: work is work

[31/03/2011 9:18:55 a.m.] Client: cricket is cricket

[31/03/2011 9:19:02 a.m.] Offshore Supplier: YES SIR

[31/03/2011 9:19:07 a.m.] Client: We are too much behind schedule Offshore Supplier

[31/03/2011 9:19:10 a.m.] Client: would appreciate it if we can get back on track please.

[31/03/2011 9:19:29 a.m.] Offshore Supplier: yes sir ..I agree

### B-4. Conversation related to Payment & Billing

[25/02/2011 2:59:00 p.m.] Client: do you anythig else from me Offshore Supplier?

[25/02/2011 2:59:45 p.m.] Offshore Supplier: what abt payments

[25/02/2011 2:59:56 p.m.] Offshore Supplier: possible for you to make some today?

[25/02/2011 3:00:04 p.m.] Offshore Supplier: yes, how much do you want me to make it for?

[25/02/2011 3:00:11 p.m.] Offshore Supplier: your wish

[25/02/2011 3:00:22 p.m.] Client: How about $1500 today?

[25/02/2011 3:00:32 p.m.] Client: that takes care of Project C / Project D

[25/02/2011 3:00:43 p.m.] Client: ($650)

[25/02/2011 3:00:45 p.m.] Client: actually

[25/02/2011 3:00:47 p.m.] Client: make it $1650

[25/02/2011 3:00:52 p.m.] Client: so I can pay you another 1k for Project A

[25/02/2011 3:02:42 p.m.] Offshore Supplier: hmm

[25/02/2011 3:02:57 p.m.] Offshore Supplier: ok

[25/02/2011 3:03:15 p.m.] Client: I think I paid $1350 last time for Project A right?

[25/02/2011 3:03:19 p.m.] Offshore Supplier: lets do this send me 500 via mass pay and lets see what happends

[25/02/2011 3:03:25 p.m.] Client: + $1k

[25/02/2011 3:03:34 p.m.] Client: so final balance remaining will be $1k + bonus

[25/02/2011 3:04:12 p.m.] Offshore Supplier: hmm....need to check

[25/02/2011 3:04:20 p.m.] Client: BRB

[25/02/2011 3:04:21 p.m.] Client: I need to check as well

[25/02/2011 3:04:26 p.m.] Offshore Supplier: ok

[25/02/2011 3:13:40 p.m.] Client: $2300 paid last time Offshore Supplier

[25/02/2011 3:13:57 p.m.] Offshore Supplier: yes

[25/02/2011 3:14:27 p.m.] Client: Which Included first installment of 1250 USD out of 3200USD for AAT

[25/02/2011 3:14:47 p.m.] Client: The remaining amount was for Project B

[25/02/2011 3:14:56 p.m.] Client: so, there is currently $1950 outstanding for Project A

[25/02/2011 3:19:31 p.m.] Client: so, $400 for Project C + $200 for Project D = $600 + $950 for Project A = $1550

[25/02/2011 3:19:39 p.m.] Client: is that ok?

[25/02/2011 3:19:59 p.m.] Offshore Supplier: i though 450 for Project C :)

[25/02/2011 3:20:15 p.m.] Offshore Supplier: 220 Project D

[25/02/2011 3:20:28 p.m.] Client: $50 / $20 was bonus

[25/02/2011 3:20:31 p.m.] Offshore Supplier: yes

[25/02/2011 3:20:33 p.m.] Client: project is not complete :)

[25/02/2011 3:20:38 p.m.] Offshore Supplier: ok ..cool

[25/02/2011 3:20:43 p.m.] Offshore Supplier: so here is what i want you to do

[25/02/2011 3:21:15 p.m.] Offshore Supplier: plz send me 500USD to my ABC paypal account via mass pay system

[25/02/2011 3:21:30 p.m.] Offshore Supplier: once I receive that

[25/02/2011 3:21:38 p.m.] Offshore Supplier: I will come to know abt the tax deduction

[25/02/2011 3:21:39 p.m.] Offshore Supplier: s

[25/02/2011 3:21:53 p.m.] Offshore Supplier: there reaming amt can come later

[25/02/2011 3:22:09 p.m.] Offshore Supplier: are you ok with that

[25/02/2011 3:22:15 p.m.] Client: Can you send me an invoice for Project C & Project D for $60 please?

[25/02/2011 3:22:17 p.m.] Client: 600*

[25/02/2011 3:22:25 p.m.] Offshore Supplier: yes

[25/02/2011 3:22:26 p.m.] Client: I will pay for $600 to ABC via mass pay

[25/02/2011 3:22:32 p.m.] Client: that covers those two projects

[25/02/2011 3:22:38 p.m.] Offshore Supplier: will draft and send right away

[25/02/2011 3:22:46 p.m.] Client: if you are happy with tax rate, I will send the remaining $950 for AAT separately

[25/02/2011 3:22:50 p.m.] Client: how does that sound?

[25/02/2011 3:22:55 p.m.] Offshore Supplier: cool!